# Title : Lattice contraction induced by resistive switching in chromium-doped $V_2O_3$: a hallmark of Mott physics

**Authors:** D. Babich[1], J. Tranchant[1], C. Adda[1,5], B. Corraze[1], M.-P. Besland[1], P. Warnike[4], D. Bedau[2], P. Bertoncini[1], J.-Y. Mevellec[1], B. Humbert[1], J. Rupp[3], T. Hennen[3], D. Wouters[3], R. Llopis[5], L. Cario[1,*], E. Janod[1,*].

[1] Université de Nantes, CNRS, Institut des Matériaux Jean Rouxel, IMN, F-44000 Nantes, France.

[2] Western Digital Co, HGST, San Jose, CA 95135 USA.

[3] Institut für Werkstoffe der Elektrotechnik II and Jülich-Aachen Research Alliance on Fundamentals of Information Technology (JARA-FIT), RWTH Aachen University, D-52056 Aachen, Germany.

[4] Paul Scherrer Inst, CH-5232 Villigen, Switzerland.

[5] CIC nanoGUNE, Tolosa Hiribidea 76, 20018 Donostia-San Sebastian, Spain

*Correspondence to: Laurent.Cario@cnrs-imn.fr, etienne.janod@cnrs-imn.fr.

**Abstract:** Since the beginnings of the electronic age, a quest for ever faster and smaller switches has been initiated, since this element is ubiquitous and foundational in any electronic circuit to regulate the flow of current. Mott insulators are promising candidates to meet this need as they undergo extremely fast resistive switching under electric field. However the mechanism of this transition is still under debate. Our spatially-resolved µ-XRD imaging experiments carried out on the prototypal Mott insulator $(V_{0.95}Cr_{0.05})_2O_3$ show that the resistive switching is associated with the creation of a conducting filamentary path consisting in an isostructural compressed phase without any chemical nor symmetry change. This clearly evidences that the resistive switching mechanism is inherited from the bandwidth-controlled Mott transition. This discovery might hence ease the development of a new branch of electronics dubbed Mottronics.

**Introduction:**

Much of the digital revolution that humanity is currently experiencing is due to the amazing development of a fundamental building block, the transistor. All key electronic components such as processors, volatile or non-volatile memories, running everyday objects like smartphones or computers are indeed based on the control of the conducting *vs*. insulator state of the transistor. However, the dimensional downscaling of this building block is now approaching fundamental limits giving an end to the Moore's law and to traditional means of development of the microelectronics industry. As a result, new concepts for system architecture [1] and new materials for information processing and memory are being explored to extend the historical pace of progress observed over the last half-century. One of the proposed solutions is the Mottronics, *i.e.* the use of Mott insulators in microelectronic devices[2,3,4]. These materials were among the most studied ones during the last decades owing to their outstanding electronic properties[5,6]. However, an insulator to metal transition (IMT) driven by a control parameter compatible with microelectronics, *i.e.* the electrical pulse, has only been demonstrated recently in Mott insulators. This property called Electric Mott Transition enabled the realization of breakthrough microelectronic devices, such as a Mott artificial neuron [7], a key building block for unconventional computation, or a new class of non-volatile memories and selectors[2,8], and thus gives hope in the realm of the Mottronics. But despite some significant advances[9,10,11], some controversies exist concerning the nature of the metallic phase which persists after the application of the electric pulse and the exact role of the electronic correlation in the transition [12].

Here we study the electric Mott transition in the emblematic Mott insulator $(V_{1-x}Cr_x)_2O_3$ and show that the main recorded difference between the filamentary metallic phase and the insulating pristine material is a lattice contraction. Combining electronic transport, conducting-atomic force microscopy, Raman spectroscopy and µ-X-ray diffraction measurements, we bring out the strong analogy between the compressed phase within the filament and the metallic phase stable under pressure beyond the Mott insulator to metal transition[13]. The mechanism of the Electric Mott transition seems therefore deeply related to the physics of the Mott insulator to metal transition.

Modern band structure theories of Mott insulators predict the occurrence of two types of insulator to metal transitions (IMT), induced either through a change of the electronic filling or of the bandwidth, with practical near equilibrium implementations with charge doping and pressure application[14,15]. Studies dedicated to out-of-equilibrium IMT's in correlated systems are more recent. The first evidence of insulator to metal transitions induced by electric field in Mott insulators dates back to the early 2000s [16] and the same phenomenology was found afterwards in numerous other narrow gap Mott insulators, including the prototypal Mott insulator systems $(V_{1-x}Cr_x)_2O_3$ [17]. In all these compounds, an electric field above a threshold value of a few kV/cm is able to break the Mott insulating state and to induce a resistive switching. Several mechanisms have been proposed based on theoretical or experimental studies to explain the destabilization of the Mott state by electric field[16,18,19], including Zener [12,20] or electronic avalanche effect [9,10]. Numerous experimental and theoretical studies have also revealed that this resistive switching is related to the formation of a metastable filamentary conducting path [9,17,21,22]. The stability in time of this filament depends on the electric pulse applied to the system. For pulses just above the threshold electric field, the filamentary path fully relaxes and disappears after the pulse, which leads to a volatile transition, while higher electric fields stabilize the filamentary path, inducing thus a non-volatile transition.

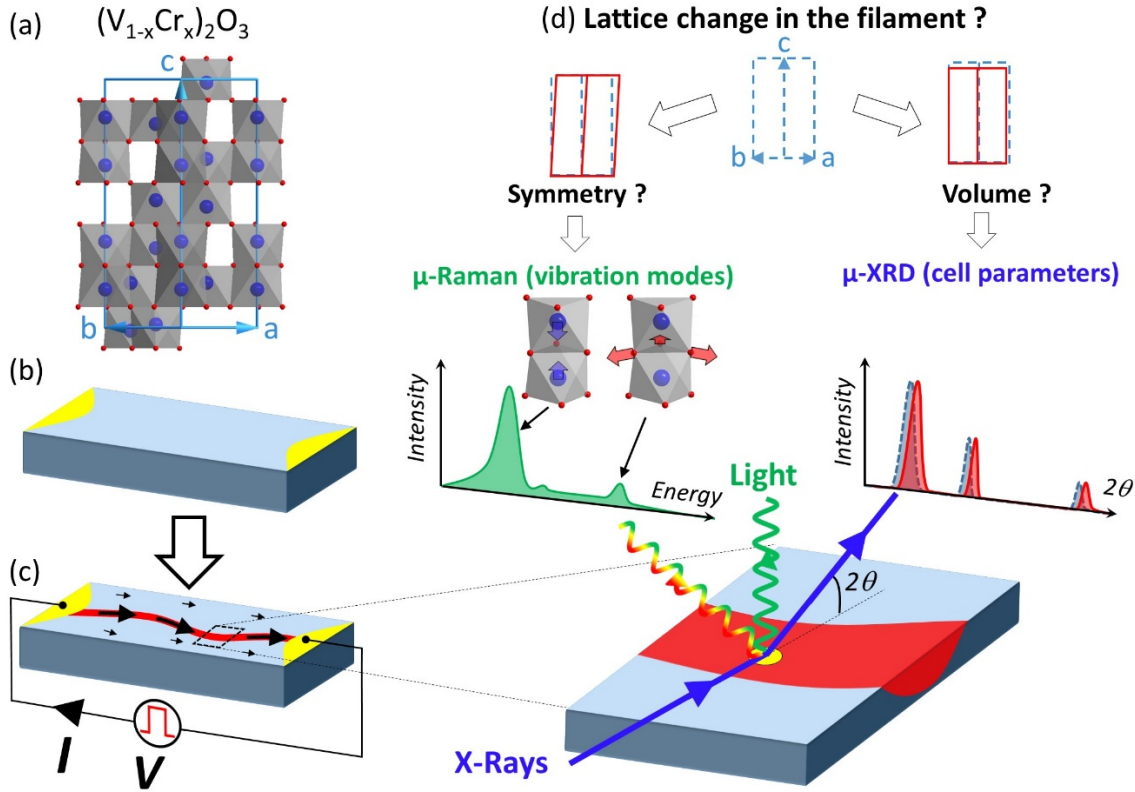

**Figure 1**: overview of techniques used in this study to probe the electronic and structural consequences of the Electric Mott Transition (EMT) in Cr-doped $V_2O_3$. (a) View of the corundum structure of $(V_{1-x}Cr_x)_2O_3$ in the paramagnetic Mott insulator (PI) phase. (b-c) Schematic drawings of the samples (either a single crystal or a 150 nm thick thin film) patterned with electrodes (in yellow) before (b) and after (c) a non-volatile Electric Mott Transition. The path created after a non-volatile EMT, represented in red, is a conducting filament with modified electronic and structural properties. (d) In order to clarify the evolution (in particular a possible symmetry-breaking or a volume change) with respect to the pristine phase, the EMT-induced metallic filamentary phase was studied by μ-Raman and μ-X-Ray diffraction mapping.

Nevertheless, important questions remain unresolved. In particular, what is the nature of the metastable metallic phase induced by the Electric Mott Transition? Is the filamentary metallic phase qualitatively different from the one induced by the classical mechanisms, *i.e.* the bandwidth- and filling-controlled IMT's ? Addressing these issues is of great fundamental interest as it could unravel an as yet unsolved basic property of highly correlated compounds placed out-of-equilibrium. But it could also facilitate the use of Electric Mott transitions in Mottronics devices thanks to a better control of this property. **Figure 1** summarizes the approach used to clarify the nature of the electric-field-induced filamentary metallic phase. We studied the macroscopic electronic response, but also the concomitant lattice response of an emblematic Mott insulator, the chromium-doped vanadium sesquioxide $(V_{1-x}Cr_x)_2O_3$. We mapped single crystals and patterned thin films using two techniques spatially resolved at the micrometer scale and able to detect possible symmetry breakings and / or volume changes, namely Raman spectroscopy and X-ray diffraction.

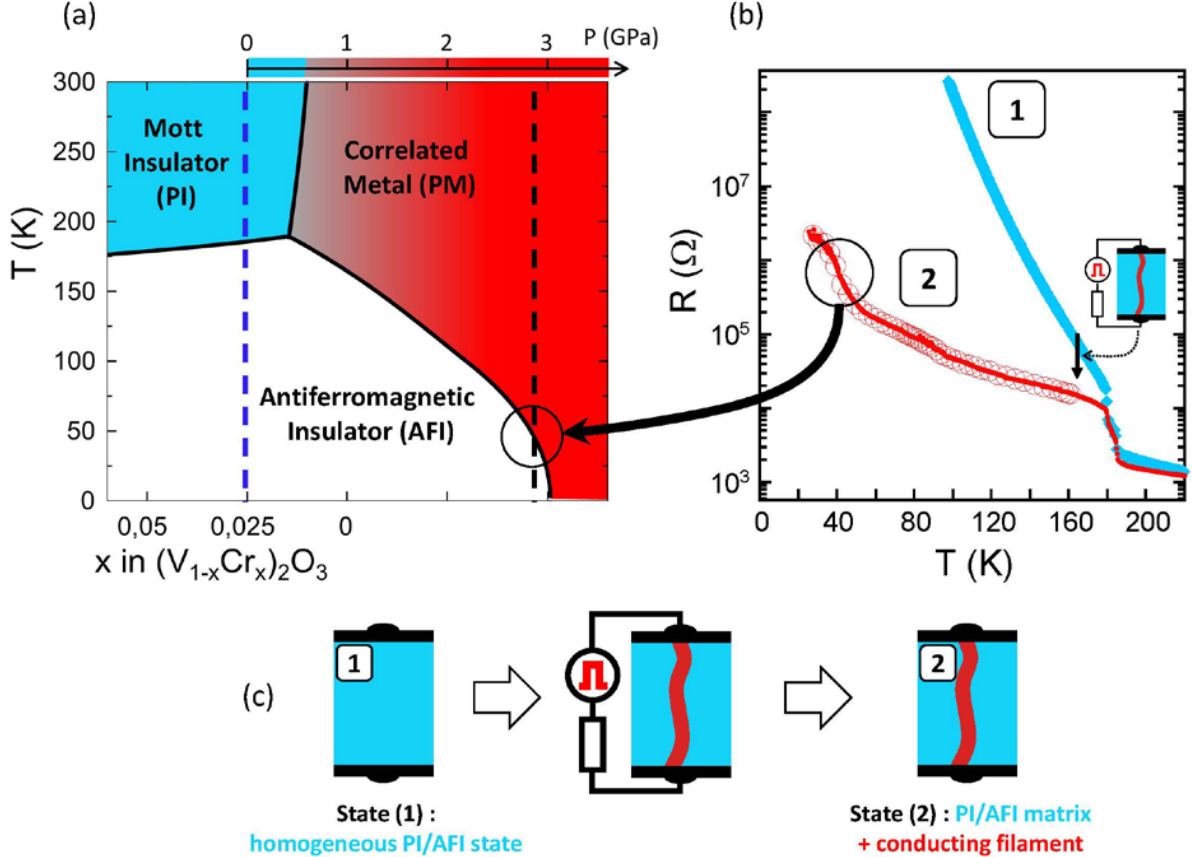

**Figure 2**: (a) temperature – composition x – pressure phase diagram of $(V_{1-x}Cr_x)_2O_3$. The pressure scale is positioned for the composition x = 0.025, using the equivalence of 0.4 GPa per Cr percent proposed in Ref. [23]. (b) Electrical resistance vs temperature of a $(V_{0.975}Cr_{0.025})_2O_3$ single crystal in its initial insulating state (1) and after application of an electric pulse inducing the creation of a conducting filamentary path (2), as schematically represented in (c).

## Results

The rich phase diagram of the prototypal Mott insulator $(V_{1-x}Cr_x)_2O_3$ reproduced in **Figure 2**-a was established thanks to the pioneering work of McWhan *et al.* at Bell Labs fifty years ago [13]. Depending on chromium content (x), pressure (P) and temperature (T), three phases appear, the Paramagnetic (Mott) Insulator (PI), the Anti-Ferromagnetic (Mott) Insulator (AFI) and the Paramagnetic Metal (PM) phase. For a single crystal with chemical composition $(V_{0.975}Cr_{0.025})_2O_3$, a PI state is expected at room temperature, with a transition to an AFI state below 180 K. **Figure 2**-b presents the main results of our electronic transport study conducted on a $(V_{0.975}Cr_{0.025})_2O_3$ single crystal. The resistance *R* vs. temperature *T* dependence in the pristine state (curve #1) is in perfect agreement with the phase diagram (Figure 2-a), with a semiconducting-like behavior ($dR/dT < 0$) in the whole studied temperature range. Moreover a ten-fold increase of resistance appears at the transition between the small gap insulator (PI phase) and the large insulator (AFI phase) at $T_{PI \to AFI}$=180 K, as expected for a composition $(V_{0.975}Cr_{0.025})_2O_3$.[24] This sample was then submitted to a 5µs electric pulse (75 V ≡ 18 kV/cm) at 160 K, leading to a non-volatile Electric Mott Transition. The subsequent *R(T)* curve measured upon cooling (see curve #2 in **Figure 2**-b) reveals a substantial drop of electrical resistance and an unexpected feature at low temperature: a broad but well-defined increase of resistance around 40 K. The resistance drop may be easily explained by a simple model proposed for other Mott insulators [17,21] considering the creation of a conducting filament embedded in the pristine material and that percolates between the contact electrodes (see **Figure 2**-c). In this model, the conductance of the filament becomes predominant at low temperature [13,25] and the additional resistance jump at 40 K in $(V_{1-x}Cr_x)_2O_3$ (see curve #2) hence originates from a transition within the conductive filament. This gives some interesting clues about the nature of the metallic phase created after the electric pulse. According to the phase diagram shown in **Figure 2**-a, a resistivity jump at 40 K is indeed expected at the PM – AFI transition of $(V_{0.975}Cr_{0.025})_2O_3$ put under a pressure of ≈ 3 GPa. This macroscopic transport study, presented more in details in Supplementary Materials, suggests therefore that the electric pulse has created a filamentary path made of the same phase, but with a compressive strain strong enough to push it in the correlated metal phase (PM) at room temperature.

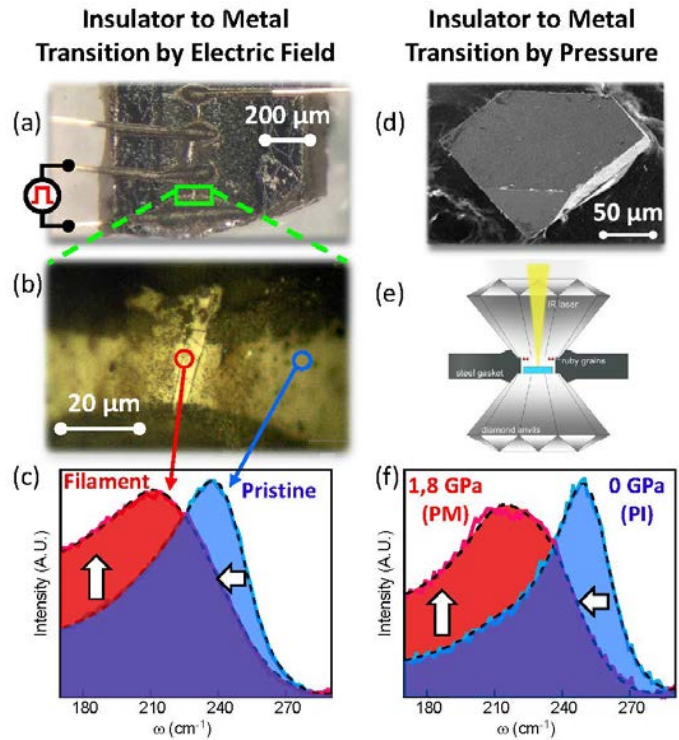

**Figure 3**: comparison of Raman spectra of the metallic states obtained in $(V_{1-x}Cr_x)_2O_3$ by application of an electric field (left, x = 0.025) and of hydrostatic pressure (right, x = 0.038). (a) Image of the $(V_{0.975}Cr_{0.025})_2O_3$ crystal, with four electrodes made of gold wires and carbon paste. Electric pulses were applied between neighboring electrodes, resulting in the creation of conducting filamentary paths. (b) Zoom on the region close to a filamentary path. (c) Raman spectra of the $(V_{0.975}Cr_{0.025})_2O_3$ crystal obtained at room temperature inside and outside the 4 µm wide conducting filament shown in (b). (d) Scanning Electron Microscope image of the $(V_{0.962}Cr_{0.038})_2O_3$ single crystal used for Raman scattering experiments under pressure. (e) Schematics of the Diamond Anvil Cell used for Raman scattering under pressure. (f) Raman spectra of a $(V_{0.962}Cr_{0.038})_2O_3$ crystal measured at room temperature in the Diamond Anvil Cell close to ambient pressure in the Mott insulator states (PI) and at 1.8 GPa in the metallic state (PM, see **Figure 2**). Details about the fit of Raman spectra (dotted lines in (c) and (f)) are presented in Supplementary Materials.

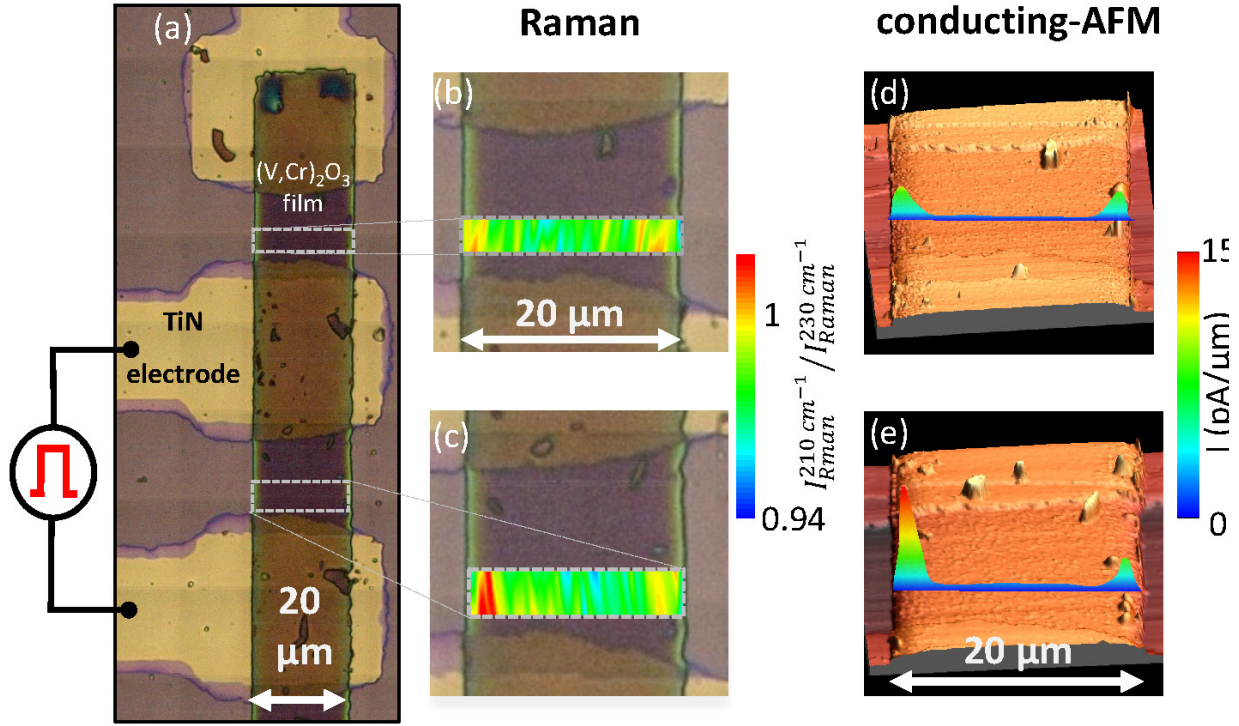

**Figure 4**: Characterization of pristine and transited thin-film device by µ-Raman and conducting-AFM. (a) General image of 150 nm thick polycrystalline $(V_{0.95}Cr_{0.05})_2O_3$ thin film with a rectangular 170 x 20 µm shape deposited on a $SiO_2$/Si substrate patterned with TiN metallic electrodes. A schematic illustration of electrodes connected to the electric pulse generator is shown. (b-c) Mapping of Raman intensities ratio measured at 210 and 230 cm$^{-1}$ in the inter-electrode domains in the pristine (b) and transited (c) cases. (d - e) Conducting-AFM mapping of the same regions shown in (b-c). All Raman and conducting-AFM measurements shown in (b-e) were made at room temperature.

To confirm this lattice contraction, a direct visualization and study at the local scale of the crystallographic structure within the filament is required. A second crystal of composition $(V_{0.975}Cr_{0.025})_2O_3$ issued from the same batch was hence subjected to a non-volatile Electric Mott Transition. **Figure 3**a-b shows that this transition leads to the formation of a 4 µm wide filamentary path bridging the two 25 µm apart conductive electrodes. The size of this filamentary structure allows to use Raman scattering to probe its crystallographic structure, thanks to the micrometric spatial resolution of this technique. The Raman spectra measured with a laser beam size of 1-2 µm both inside and outside the filament are shown in **Figure 3**–c and in **Figure S-6** and **S-7**. Both spectra show only the usual signatures of the corundum structure of the $(V_{1-x}Cr_x)_2O_3$ system [26,27], *i.e.* a broad structure located between 180 and 280 cm$^{-1}$ and two peaks around 300 and 500 cm$^{-1}$. No additional peaks are detected within the filament that could be associated either to a symmetry-breaking or to the formation of other $VO_x$ phases. However, the $A_{1g}$ Raman peak around 240-250 cm$^{-1}$ measured in the filament is shifted by more than -20 cm$^{-1}$ compared to the pristine Raman spectrum (see **Figure 3**-c). Moreover, a significant increase of the Raman signal is observed below 200 cm$^{-1}$. These modifications of Raman spectra, discussed in more details in Supplementary Materials, support the existence of electronic and structural changes within the filament. To check if they could be assigned to a lattice contraction, a Raman study under pressure was then carried out at room temperature on a $(V_{0.962}Cr_{0.038})_2O_3$ single crystal. For this composition, a pressure-induced Mott insulator-to-metal transition is observed for a pressure above 1.2 GPa. The Raman spectra measured under 1.8 GPa and at ambient pressure, displayed in **Figure 3**-f, disclose that this Mott transition causes both a shift of the *$A_{1g}$* Raman vibration mode and an increase of Raman intensity below 200 cm$^{-1}$. Both changes are similar to the one observed between the pristine material and the core of the filamentary path created by the Electric Mott transition. Overall, this micro-Raman study provides additional evidences that the Electric Mott Transition yields a compressed, isostructural and metallic $(V_{1-x}Cr_x)_2O_3$ filamentary phase. It strongly supports the lattice contraction scenario presented above based on resistivity measurements.

To further confirm the contraction of the lattice within the filament, a spatially resolved X-ray diffraction experiment was engaged using a 1 µm beam size at the micro-XAS beamline of the Swiss Light Source synchrotron facility. The penetration depth of X-ray being far bigger than the thickness of the filamentary path, we worked on polycrystalline thin films to maximize the signal coming from the filament over that of the matrix. In that purpose, a 150 nm polycrystalline $(V_{0.95}Cr_{0.05})_2O_3$ thin film with a rectangular 170 x 20 µm shape was deposited on a SiO$_2$/Si substrate patterned with TiN metallic electrodes (see **Figure 4** and **Figure S4**). Details about the preparation method of this device are given in Supplementary Materials. A 1 µs / 70 V electric pulse (≡ electric field $E \approx 35$ kV/cm) was applied at 160 K between two TiN electrodes to induce a non-volatile Electric Mott Transition. The sample was then characterized by conducting-atomic force microscopy (c-AFM) and by Raman spectroscopy. Atomic Force Microscopy (AFM) images shown in **Figure 4**-(d-e) reveal that the inter-electrode domain of the $(V_{0.95}Cr_{0.05})_2O_3$ film affected by the Electric Mott Transition does not display any morphological change. **Figure 4**-(d-e) also shows the distribution of current measured by applying a 25 mV bias between the conductive AFM tip and the neighboring electrode, integrated over similar areas in regions affected by the EMT (**Figure 4**-e) and in pristine regions (**Figure 4**-d). Interestingly, the map shown in **Figure 4**-e unveils a clear ≈ 2 µm wide zone of higher conductivity close to the left edge of the $(V_{0.95}Cr_{0.05})_2O_3$ stripe. **Figure 4**-c shows that the Raman mapping measured in the highly conducting region uncovered by c-AFM presents a tendency towards metallicity (ratio of Raman intensities between 210 and 230 cm$^{-1}$ > 1, see **Figure 3**) compared to spectra measured outside of this zone. The c-AFM and Raman measurements confirm therefore the presence of a metallic filamentary path between the two inner electrodes of the polycrystalline device with features similar with those observed on the crystal.

A µ-XRD experiment was then engaged by mapping a 200 x 45 µm zone containing the whole patterned $(V_{0.95}Cr_{0.05})_2O_3$ stripe of the device (170 x 20 µm). Details about the µ-XRD experiments are provided in Supplementary Materials. The map consists of 9191 (101 x 91) "pixels" of 5 x 1 µm, each of them corresponding to a 2D powder X-Ray diffractogram (see **Figure 5**-a). An azimuthal integration of the Debye rings was performed for each 2D powder X-Ray diffractogram to extract the intensity *vs 2θ* plots. Representative examples of the integrated diffraction patterns outside and inside the filament are shown in **Figure 5**-b. Outside the filament the diffraction pattern exhibits the expected Bragg peaks (104), (110) and (113) of the corundum structure. Inside the filament the same peaks are observed, but slightly shifted. Moreover, the absence of new diffraction peak discards the change of crystallographic symmetry or the formation of a chemical phase with different composition. The mechanism proposed in Ref. [28], which suggests the creation of a $V_5O_9$ phase to explain the resistivity jumps in the low temperature phase, can therefore be excluded in our case. This crude comparison of XRD patterns in **Figure 5**-b already confirms that the $(V_{0.95}Cr_{0.05})_2O_3$ phase within the filament keeps the corundum structure but with a different cell volume. The *a* and *c* unit cell parameters were then refined for each pixel based on the position of the (104), (110) and (113) Bragg peaks and were represented as maps in **Figure 5**-c.

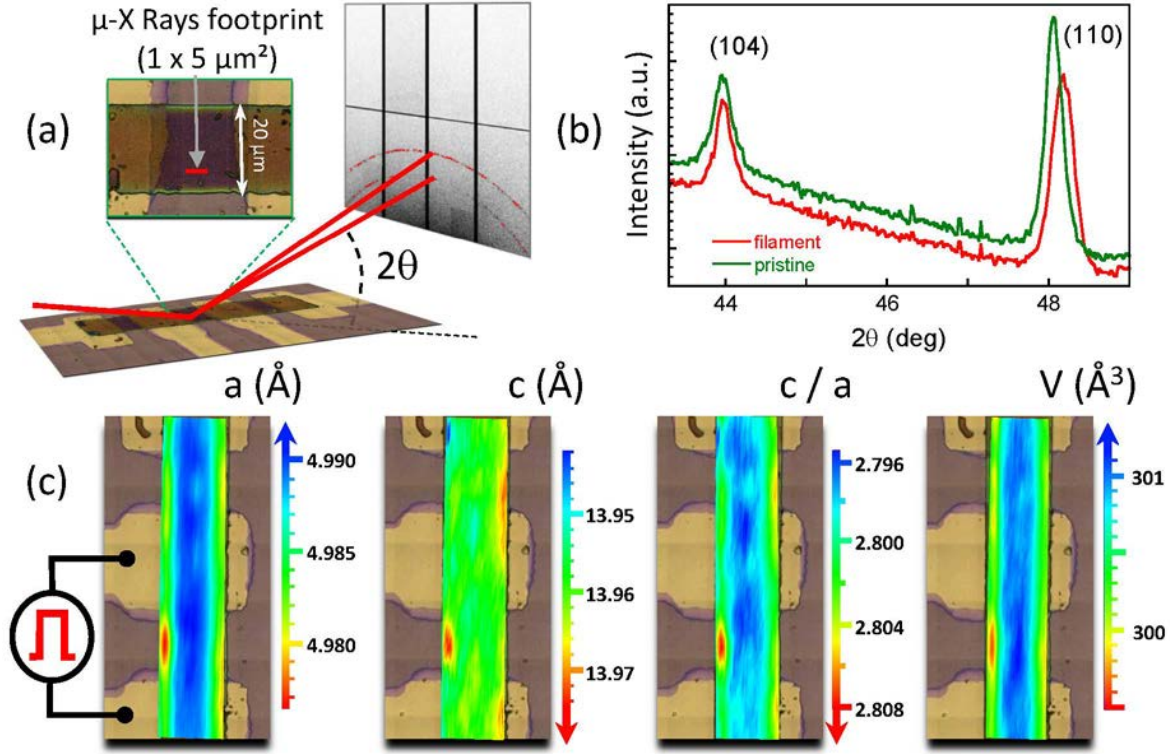

**Figure 5** : µ-XRD mapping experiment performed at room temperature on a $(V_{0.95}Cr_{0.05})_2O_3$ thin-film device submitted to a Electric Mott Transition. (a) Schematic view of the µ-XRD geometry, with a fixed incident angle of 11° between the X-Ray beam and the film, leading to an X-ray beam footprint of 1 µm wide and 5 µm long. A Dectris Eiger 4M 2D detector was used to register the Debye diffraction rings. The mapping was obtained by moving the sample holder with steps of 0.5 and 2.5 µm along two directions. (b) Comparison of µ-XRD diffractograms inside and outside the filamentary path. (c) Maps of lattice parameters *a* and *c*, as well as of the *c / a* ratio and the hexagonal unit cell volume extracted from XRD refinements of the 9191 measured diffractograms.

These maps reveal a remarkable feature: in the filamentary region bridging the two central electrodes, an increase of the *c* parameter (from 13.95 to 13.97 Å, ≈ + 0.15%) with respect to the rest of the sample is clearly observed. Conversely, the unit cell parameter *a* is contracted by about -0.4% compared to the normal zones. Overall the unit cell volume within this filament undergoes a contraction of -0.6% compared to the pristine material. These structural evolutions within the filament are particularly interesting, since they allow discarding two trivial scenarios. First, a filament's creation due exclusively to Joule heating followed by a thermal quench is not compatible with our data, since it would result in an *increase* of the unit cell parameter *a* and of the unit cell volume *V*. Second, our data do not point towards the creation, within the filament, of a non-stoichiometric phase with a (V+Cr) : O ratio departing from 2:3. Indeed such non-stoichiometry induces a concomitant decrease of both *a* and *c* parameters[29,30] (see discussion about non stoichiometry in $V_2O_3$ provided in Supplementary Materials), which contrasts with the observed increase of the *c* parameter measured in the filamentary path shown in **Figure 5**-c. Conversely, the very specific observed behavior (increase of the unit cell *c*, and decrease of both the unit cell *a* and volume *V*) is strikingly similar with changes observed at the isostructural Mott transition driven by hydrostatic pressure [13,31] shown in **Figure S2**. A + 0.26 % increase of *c* parameter and a decrease of - 0.65 % (-1.04 %) decrease of *a* parameter (volume *V*) is indeed observed at the pressure induced Mott transition at room temperature for a $(V_{0.972}Cr_{0.028})_2O_3$ single crystal [31]. Finally we note that the creation of the conducting filamentary path is a reversible process, since we have previously demonstrate in Refs. [32,33] the possibility to switch back and forth between high and low resistance states by applying electric pulses of the same polarity.

The electronic transport and c-AFM measurements, as well as the µ-Raman and µ-XRD studies, presented in this work are hence consistent with the creation by electric field of a metallic filamentary phase whose unit cell volume is contracted compared to that of pristine $(V_{1-x}Cr_x)_2O_3$ Mott insulator. Despite stimuli of seemingly very different nature, the electric-field-induced contraction resembles the one observed under pressure at the transition between the Mott insulator (PI) and paramagnetic metal (PM) phases. Interestingly, the pressure-induced volume contraction at the Mott transition can be rationalized within the compressible Hubbard model [34,35]. In this approach, the lattice degrees of freedom react to the softening of electron degrees of freedom close to the Mott transition, resulting in a volume change of the lattice in order to avoid the unphysical situation of negative bulk compressibility[i]. The results presented in this paper strongly suggest that a similar mechanism is at play at the Electric Mott transition, *i.e.* that the creation of excited and delocalized electrons during the electric pulses leads to a compressive lattice response. Overall, this work provides strong indication that Mott physics plays a pivotal role in the insulator to metal transition observed under electric field in $(V_{1-x}Cr_x)_2O_3$. Our study therefore suggests that the relevant theoretical approach to fully capture the Electric Mott Transition is the out-of-equilibrium compressible Hubbard model, which, to the best of our knowledge, has never been studied until now. Finally, the direct evidence of a lattice compression at the Electric Mott Transition in $(V_{1-x}Cr_x)_2O_3$ shown in this work might rationalize *a posteriori* the presence of electric-field-induced superconductivity in the canonical Mott insulator $GaTa_4Se_8$ [37,38], which is also superconducting under pressure [39]. This suggests that the lattice compression at the Electric Mott Transition could be a universal property of Mott insulators. Overall, the advances in understanding the mechanism of the Electric Mott Transition gained in this study will greatly facilitate the implementation of advanced Mottronics applications such as Resistive Mott memories and neuromorphic components in the near future [7,40].

**Data availability**

The datasets generated during and/or analysed during the current study are available from the corresponding authors on reasonable request.

---

[i] A similar situation occurs at the first order liquid – gas phase transition, where the volume change at the transition also occurs to avoid an unphysical negative bulk modulus. The equivalence between the Mott IMT and the liquid-gas transition is discussed in more details in ref. [36].

**Acknowledgments.**

The authors thank the University of Nantes for funding the PhD of D. B. in the framework of the French-Japanese International Associated Laboratory "Impacting Materials by Light and Electric pulses, and watching real-time Dynamics" (IM-LED). We thank Zohra Khaldi for her help in the preparation of single crystals. This work was partly supported the ANR Elastica, grant ANR-16-CE30-0018). The research leading to these results has received funding from the European 2020 research and innovation program under grant agreement n°730872 project CALIPSOplus.

**Author contributions:**

D. Babich : investigation, data curation, formal analysis, resources, validation ; J. Tranchant : conceptualization, investigation, **writing – review & editing** ; C. Adda : investigation, resources; B. Corraze : conceptualization, investigation, supervision, **writing – review & editing** ; M.-P. Besland : investigation, supervision ; P. Warnike : investigation, writing – **review & editing** ; D. Bedau : conceptualization, investigation, writing – **review & editing** ; P. Bertoncini : investigation, writing – **review & editing** ; J. - Y Mevellec : methodology, writing – **review & editing ;** B. Humbert : methodology ; J. Rupp : investigation, writing – **review & editing ;** T. Hennen : investigation, writing – **review & editing** ; D. Wouters : supervision ; R. Llopis : resources ; L. Cario : conceptualization, supervision, **writing – original draft, writing – review & editing;** E. Janod : conceptualization, investigation, supervision, **writing – original draft**, **writing – review & editing**.

**Competing interests**

Authors declare no competing interests.

# Supplementary Materials for

# Lattice contraction induced by resistive switching in chromium-doped V$_2$O$_3$: a hallmark of Mott physic$_s$ r

Correspondence to: laurent.cario@cnrs-imn.fr, etienne.janod@cnrs-imn.fr

**This PDF file includes:**

- Materials and Methods
  - ✓ Details on experimental techniques
  - ✓ Samples and device preparation

- Supplementary Text
  - ✓ Transport study on the (V$_{0.975}$Cr$_{0.025}$)$_2$O$_3$ single crystal
  - ✓ Complementary Raman spectroscopy data on (V$_{1-x}$Cr$_x$)$_2$O$_3$ single crystals
  - ✓ General discussion about the non-stoichiometry issue in the V$_2$O$_3$ system

- Figs. S1 to S7

- References

**Materials and Methods**

Details on experimental techniques

   *1. Transport and electric pulse application*

   The low-bias resistance was measured using either a Keithley 236 source-measure unit or a Keithley 6221 current source associated with a Keithley 2182A nanovoltmeter. For the electric pulses application, thin-film devices and crystals were connected in series with a load resistance $R_{load}$ playing the role of the current limiter and ranging in the 5–10% of the sample resistance. Both the crystals (see Figures 2 and 3 in the article) and the TiN pads of the device including Cr-doped $V_2O_3$ film (see Figures 4 and 5 in the article) were contacted with 17 µm gold wires using a carbon paste (Electrodag PR-406). After the required annealing in vacuum at 150 °C for 30 min, the measured contact resistances do not exceed 10-20 % of the sample resistance. Voltage pulses were applied using either a pulse generator Agilent 8114 or an arbitrary function generator Agilent 81150A.

   *2. Conducting Atomic Force Microscopy*

   Conductive – AFM (c-AFM) measurements were performed in air using the c-AFM module of the Multimode 8 nanoscope V (Bruker). Conducting platinum silicide coating tips were used (SCM-PTSI, Bruker).

   *3. Raman spectroscopy*

   Raman spectra were recorded using a microRaman Invia (Renishaw) using an excitation at 785 nm. The spectral resolution achieved with the use of gratings of 1200 grooves per millimeter was 3 cm$^{-1}$. The microscope was equipped with an automated XYZ table allowing mapping with a spatial resolution of 1µm.

   Raman scattering experiments under pressure were performed in the Membrane Diamond Anvil Cell (MDAC) schematized in
   **Figure S** 1-a. The $(V_{0.962}Cr_{0.038})_2O_3$ single crystal and a few ruby micro-crystals were placed in the pressure chamber prepared by drilling a 300 µm hole in a stainless-steel gasket (see
   **Figure S** 1-b-c). Pure methanol was used as pressure transmitting medium. The pressure was estimated by monitoring the fluorescence lines of ruby excited by a 514 nm laser. Two Raman spectra, measured on the $(V_{0.962}Cr_{0.038})_2O_3$ crystal using a 785 nm laser, are shown in
   **Figure S** 1-d. The broad contribution between 350 and 700 cm$^{-1}$ corresponds to the Raman signal of methanol, and complicates the measurement and interpretation of the $A_{1g}^{(2)}$ phonon of $(V_{0.962}Cr_{0.038})_2O_3$ around 510 cm$^{-1}$.

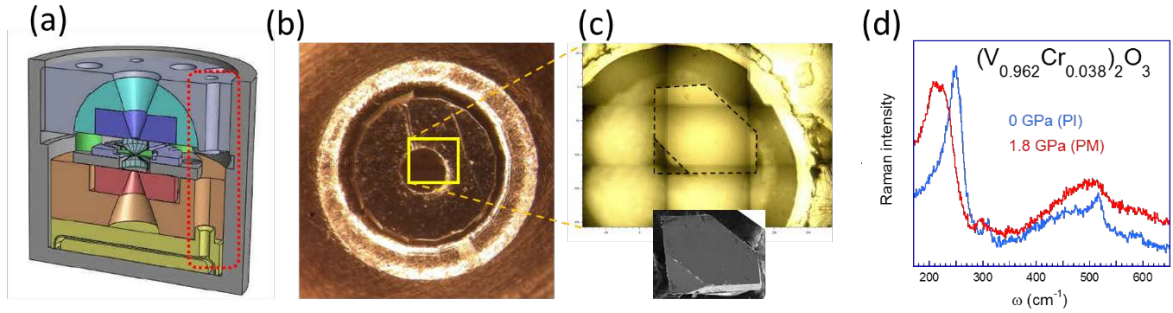

**Figure S 1 :** (a) schematic drawing of the Diamond Anvil Cell (DAC) used for Raman under pressure experiments. (b) Image of the gasket drilled with a 300 µm hole to define a pressure chamber, taken from the top of the DAC. (c) Optical image of the single $(V_{0.962}Cr_{0.038})_2O_3$ crystal inside of the pressure chamber. Inset : Scanning Electron Microscope Image of the crystal. (d) Typical Raman spectra of $(V_{0.962}Cr_{0.038})_2O_3$ measured inside of the DAC, at room pressure in the paramagnetic insulator (PI) state and at 1.8 GPa in the paramagnetic metal (PM) state.

*4. Micro-X-Rays Diffraction mapping experiments*

µ-X-ray diffraction experiments were carried out at the µ-XAS beamline (SLS Synchrotron, Switzerland). We chose a rather low X-Rays energy of 6.1 keV ($\lambda$ = 2.0325 Å) in order to keep a beam diameter as small as 1µm. The calibration of the experiment geometry (sample-detector distance, orientation of the Dectris Eiger 4M two-dimensional detector with respect to the beam) was performed with the standard $Na_2Ca_3Al_2F_{14}$ compound. An XRD experiment was performed using powders of this compound inserted into a glass capillary placed at the same position as the sample. The calibration procedure was done using the Fast Azimuthal Integration pyFAI python package[1].

In order to map the 170 x 20 µm stripe of $(V_{0.95}Cr_{0.05})_2O_3$ thin film shown in **Figure S 4.a**, 9191 two-dimensional images were recorded during the µ-XRD experiment (4 seconds per (x,y) position ➔ *i.e.* a total measurement time ≈ 10 h). We moved the sample holder with 0.5 µm steps along (101 steps) and perpendicular (91 steps) to the $(V_{0.95}Cr_{0.05})_2O_3$ stripe direction, resulting in a total map size of 250 x 45 µm. Each of the 9191 diffractograms was then azimuthally integrated to obtain a 1D "Intensity *versus* Q" curve. Then the position of the (104), (110) and (113) Bragg peaks were fitted on each diffractogram to extract the *a* and *c* unit cell parameters shown in **Figure 5** of the article.

We note that the thin film preparation process described in the next section led to an edge effect. As shown in **Figure 5** of the article, all the edges of the $(V_{0.95}Cr_{0.05})_2O_3$ stripe indeed display a slightly compressed unit cells *a* parameter and volume compared to the center of the stripe. This effect explains the increased electrical conductivity close to the edges, which appears even in the pristine state as indicated by **Figure 4.d** of the article. These edges with a slightly enhanced conductivity in the pristine state explains why the filamentary path was created close to an edge during the electric Mott Transition, as indicated in **Figures 4** and **5** of the article. However, the evolution of the *c* parameter shown in **Figure 5.c** clearly proves that there is a qualitative difference between the overall volume contraction induced by the Electric Mott Transition and that observed at the edges in the pristine state. Almost no edge effect is indeed detected for the *c* unit cell parameter in the pristine state, while a clear *increase* of *c* appears in the filamentary path created during the EMT. This clearly demonstrate that the compressive edge effect in the pristine state is too small to generate a full insulator to metal transition (IMT). Conversely, both fingerprints of the IMT, namely a drop of the *a* parameter and an increase of *c*, appears inside of the filamentary path created during the Electric Mott Transition.

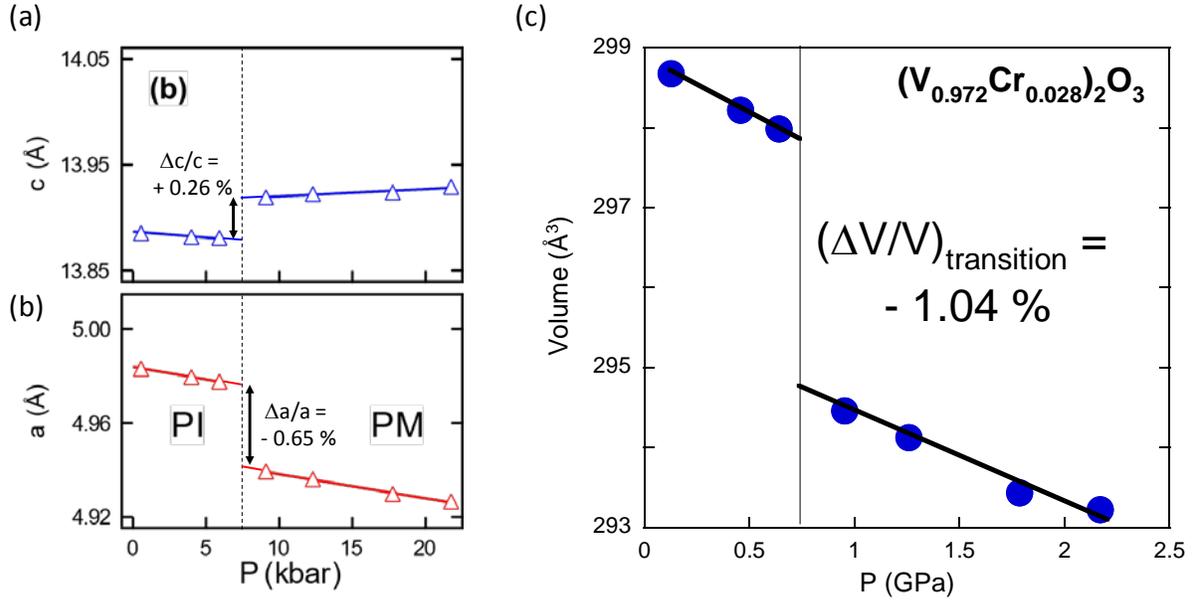

**Figure S 2** : (a-b) Evolution with pressure of the unit cell parameters *c* and *a* of a $(V_{0.972}Cr_{0.028})_2O_3$ powder, extracted from ref. 2. (c) Deduced variation of the unit cell volume with pressure.

Samples and device preparation

1. *$(V_{1-x}Cr_x)_2O_3$ single crystals : synthesis and characterization*

The $(V_{1-x}Cr_x)_2O_3$ single crystals used for transport and Raman measurements were prepared using a sulfur-assisted chemical-vapor transport method. The synthesis is performed in two steps. After a first drying at 400°C, $V_2O_5$ powder (Aldrich, >99.6%) was mixed with $Cr_2O_3$ (Prolabo, 99.9%) in the appropriate ratio. The mixture was then placed in an oven at 900°C for 10 h under a 95% Ar–5% $H_2$ gas flow. Half a gram of the single phase $(V_{1-x}Cr_x)_2O_3$ powder obtained was then introduced in a silica tube, with 40 mg of sulfur as a vapor phase transport agent. The tube was vacuum sealed, heated up to $T_{max}$ (950-1050°C) in a furnace with a temperature gradient (≈ 10°C/cm), slowly cooled down to 900°C (from - 0.5 to - 2°C/h) and finally fast cooled (−300°C/h) to room temperature. Such treatment allows obtaining small single crystals (typical size <300 µm) within the preparation. The chromium substitution rate on each individual crystal used in this work was determined by Energy-dispersive X-ray spectroscopy (EDXS) analyses carried out using a scanning electron microscope JEOL 5800.

2. *Device based on $(V_{0.95}Cr_{0.05})_2O_3$ thin film*

The patterned planar devices used in this work (see **Figures 4 and 5** of the article) were fabricated within a collaboration between the Jean Rouxel Institute of Materials (France) and the CIC nanoGUNE (Spain). The main steps of the fabrication process, depicted in **Figure S 3**, consist

in :
- step 1- Photolithography defining the electrodes: deposition of negative photoresist (nLOF 2070), exposure, development of photoresist.
- step 2-3- Etching of the TiN layer, and removal of the photoresist remaining layer.
- step 4- Photolithography defining the stripe shape of active material: deposition of positive photoresist (S1818), exposure and development.
- step 5-6- Deposition of $(V_{0.95}Cr_{0.05})_2O_3$ thin film (active material), and lift-off. A final annealing is then performed (see below).

The 150 nm thick $(V_{0.95}Cr_{0.05})_2O_3$ films were deposited by a co-sputtering technique at room temperature, and were then annealed under a controlled reducing atmosphere at 500° C for 10 h. Details about the synthesis conditions are given in ref. 3. Chemical analyses performed by EDXS confirm the presence of three elements V, Cr, and O in the films and give a Cr/(V + Cr) ratio of 0.053(4) in good agreement with the targeted chemical composition.

After the application of the electric pulses leading to the resistive switching, the samples used in this study (see **Figure 4 and 5** in the article) were characterized by Scanning Electron Microscopy (SEM) and c-AFM. It clearly appears both in **Figure S 4**-a-b (SEM) and **Figure S 4**-c (c-AFM) that the resistive switching does not result in any visible microstructural changes compared to a pristine state.

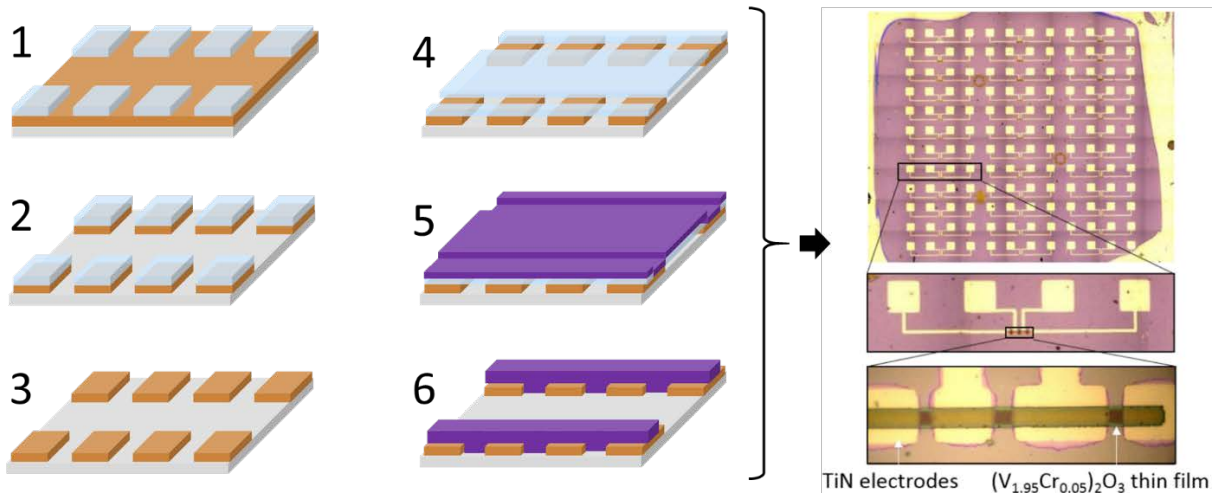

**Figure S 3** *:* 1-6 (left part): schematic fabrication process of the patterned planar devices made of titanium nitride electrodes (in brown) and $(V_{1.95}Cr_{0.05})_2O_3$ thin film (in purple) as active material. The right part corresponds to optical micrograph.

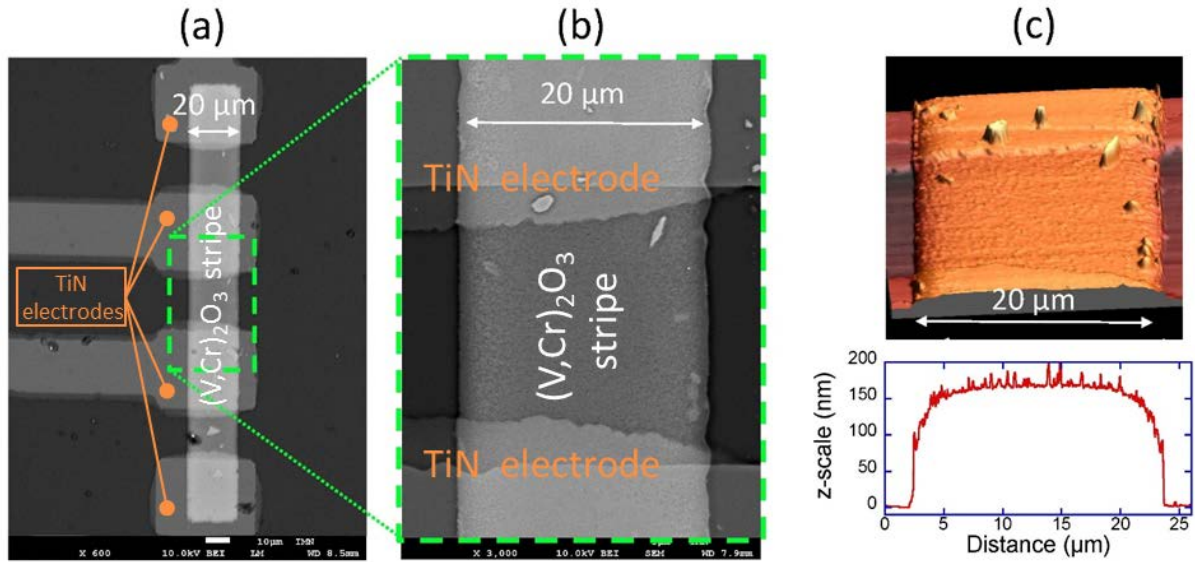

**Figure S 4** : (a-b) Scanning Electron Microscope and (c) Atomic Force Microscope (AFM) images of a device based on $(V_{0.95}Cr_{0.05})_2O_3$ thin film. The images were obtained after the application of electric pulses yielding the Electric Mott Transition (see **Figures** 4-5 in the article). The z-profile of the 170 µm long / 20 µm wide polycrystalline $(V_{0.95}Cr_{0.05})_2O_3$ stripe is represented in the lower part of (c), and indicates a film thickness of 150 nm. AFM images were analyzed using the WSxM software package[4]. The irregular edges of the TiN electrodes shown in (b) are due to an imperfect lift-off process.

**Supplementary Text**

Transport study on the $(V_{0.975}Cr_{0.025})_2O_3$ single crystal

**Figure 2** of the article discloses a part of a broader transport study carried out on a $(V_{0.975}Cr_{0.025})_2O_3$ single crystal before and after an Electric Mott Transition. The results of an extended study are depicted in **Figure S 5**. After measuring the resistance vs temperature curve in the pristine insulating state and just after the application of electric pulses, the $(V_{0.975}Cr_{0.025})_2O_3$ was then submitted to several cooling - warming cycles between 40 and 300K. In the first temperature cycle, a resistance jump appears around 40 K (red open circles, curve #2 in **Figure 2** and in **Figure S 5**). Upon heating (red filled circles, curve #2 in **Figure S 5**), the jump in the $R(T)$ curve around 40 K is essentially unmodified, with only a small hysteresis of $\approx$ 1K. At higher temperature, an attenuated $R(T)$ drop persists at the AFI $\rightarrow$ PI transition, and the resistance almost superimposes that of the pristine compound above 180 K. Surprisingly, after less than fifteen minutes spent at 290 K, the $R(T)$ curve measured subsequently by cooling again from room to low temperature (see curve #3 in **Figure S 5**) unveils a huge shift of the low temperature transition from 40 K to 105 K. Subsequent "warming to 290 K – cooling down to 40 K" cycles did no longer modify the resistance curve, with a low temperature transition maintained at $105 \pm 2$ K.

As mentioned in the article, these observations can be explained by the creation of a conducting filament which percolates between the contact electrodes. This explanation was already proposed for other Mott insulators,[5] but it has a striking consequence in $(V_{1-x}Cr_x)_2O_3$. Above the transition temperature $T_{PI \rightarrow AFI}$, the conductance of the filament is indeed low compared to that of the pristine part, letting the macroscopic resistivity almost unchanged. Conversely, the conductance of the filament becomes predominant below $T_{PI \rightarrow AFI}$ = 180K, since the conductance of the pristine part (curve #1 in **Figure S 5**) decreases sharply in the AFI phase.[6,7] As a consequence the additional resistivity jumps observed at 40 K in curves #2 and then at 105 K in curve #3 are attributed to transitions within the conductive filament. These transitions give therefore some clues about the nature of the metallic phase created after the electric pulse. According to the phase diagram of the $(V_{1-x}Cr_x)_2O_3$ system, the resistivity jumps observed at 40 K and 105 K could correspond to the PM – AFI transitions that exist at these temperatures in $(V_{0.975}Cr_{0.025})_2O_3$ put under pressure at $\approx$ 3 GPa and $\approx$ 2 GPa, respectively. Therefore, these measurements strongly suggest that the electric pulse has created a filamentary path made of the same phase, but under a compressive strain corresponding to a pressure of 2 - 3 GPa, *i.e.* strong enough to make it metallic. Within this picture, the shift of the transition temperature from 40 K to 105 K is simply explained by the strain relaxation caused by the double crossing (*i.e.* by heating and then cooling) of the AFI $\rightarrow$ PI transition in the pristine zone surrounding the filament. The abrupt changes of the *a* (+0.24 %) and *c* (-0.57 %) parameters of the pristine zone at the transition[6] might indeed relax the stress state of the filament and explain the transport properties shown in **Figure S 5**. Finally, the shift of the transition temperature observed after bringing back the sample to room temperature discards a possible change of stoichiometry within the filamentary path. At this temperature the ionic mobility is indeed far too low to induce a stoichiometry change in a measurable time.

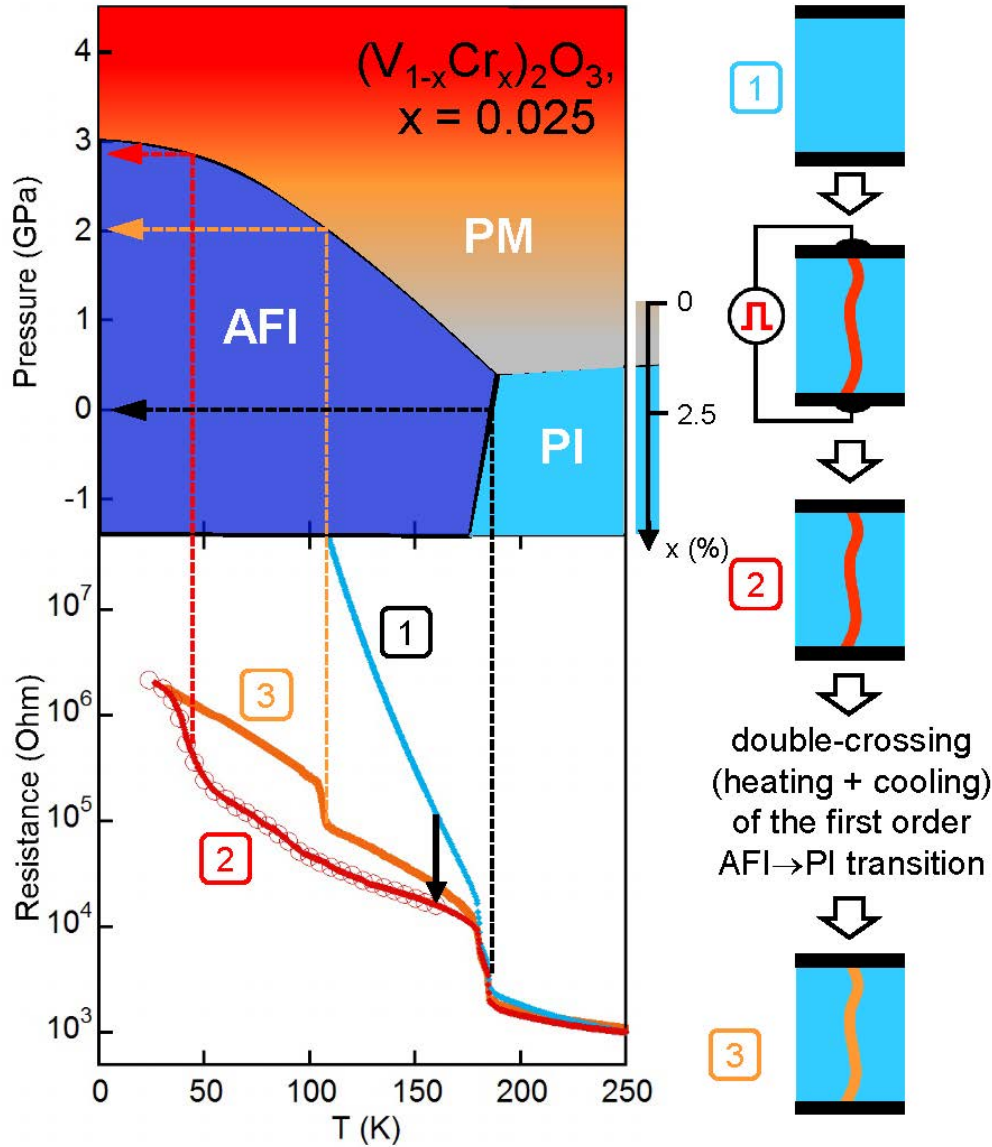

**Figure S 5** : Results of extended transport study on a $(V_{0.975}Cr_{0.025})_2O_3$ single crystal submitted to Electric Mott Transition at 160 K, in complement to Figure 2 in the article.

Complementary Raman spectroscopy data on $(V_{1-x}Cr_x)_2O_3$ single crystals

    In complement with the Raman study of the conducting filament shown in **Figure 3** of the article, we present here additional Raman data obtained on the same $(V_{0.975}Cr_{0.025})_2O_3$ crystal in the PI phase and in the PM phase under pressure, as well as on a Cr-free $V_2O_3$ crystal in the PM phase. Both crystals crystallize in the $R\bar{3}c$ corundum crystallographic structure with the $D_{3d}$ point group[8]. Seven Raman-active modes are expected, two $A_{1g}$ and five $E_g$ modes. **Figure S 6** shows that three intense Raman modes can be clearly identified in $(V_{1-x}Cr_x)_2O_3$ (x = 0.025), which can be assigned to two $A_{1g}$ and one $E_g$ modes thanks to the modelling work of Yang and Sladek[9].

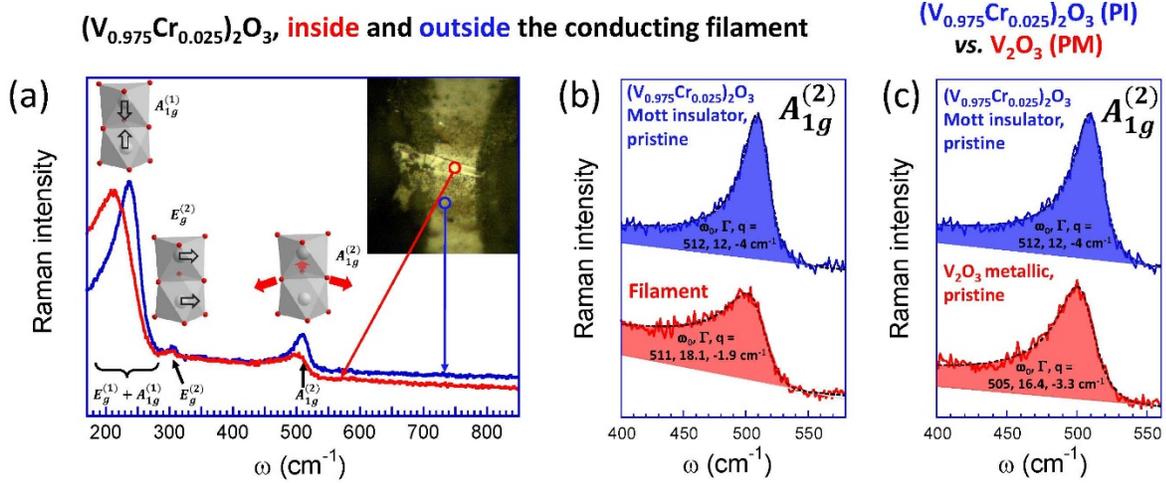

**Figure S 6 :** (a) Full Raman spectra measured inside (red curve) the conducting filament and outside in the pristine state (blue curve). The atomic displacements associated with the three most intense Raman modes are schematically represented.[9] (b) Same Raman spectra as in (a), zoomed on the $A_{1g}^{(2)}$ mode (see text). (c) Comparison of the $A_{1g}^{(2)}$ mode measured in the PI phase on a $(V_{0.975}Cr_{0.025})_2O_3$ crystal and in the PM phase on a $V_2O_3$ crystal at room temperature. The strongly asymmetric shape results from a Fano resonance, i.e. a coupling between the $A_{1g}^{(2)}$ phonons and electronic transitions within the $a_{1g}$ – bonding band[10]. Accordingly, the Raman modes were fitted according to a model appropriate to describe the Fano effect : $I(\omega) = I_0(q + \varepsilon)^2/(1 + \varepsilon^2)$, with $\varepsilon = (\omega - \omega_0)/\Gamma$, where $\omega_0$ is the resonance frequency, $\Gamma$ is a measure of the line width, and $q^{-1}$ is the Fano coupling coefficient. Clearly, $q^{-1}$ is enhanced (or equivalently $q$ is lowered) in the metallic phases with respect to the PI value, both in the standard PM phase of pure $V_2O_3$ and in the conducting filament.

In the article, **Figure 2**-c shows the changes in Raman spectra between the pristine state and the metallic state in the conducting filament in the energy window 170-290 cm-1. **Figure S 6**-a shows the same data on a larger energy window (170 - 850 cm-1). This global view shows that essentially the same Raman vibration modes appear inside and outside of the conducting filament: two $A_{1g}$ modes around 240 cm$^{-1}$ ($A_{1g}^{(1)}$) and 510 cm$^{-1}$ ($A_{1g}^{(2)}$), as well as an $E_g$ modes of weak intensity around 210 cm$^{-1}$ ($E_g^{(1)}$) and another more intense one at 305 cm$^{-1}$ ($E_g^{(2)}$). These measurements, in perfect agreement with the studies published in the system $(V_{1-x}Cr_x)_2O_3$[11,12] support the scenario of a filament made up exclusively of the phase $(V_{1-x}Cr_x)_2O_3$ phase, without any other phase of different chemical composition.

In the article, **Figure 3** shows that a softening of the $A_{1g}^{(1)}$ mode occurs in the filament, very similar to that appearing at the pressure – induced insulator transition. **Figure S 6**-b shows that the $A_{1g}^{(2)}$ mode around 510 cm$^{-1}$ presents a similar shift, but also a striking change in the peak shape of the filament compared to the pristine state. The $A_{1g}^{(2)}$ peak in the filament indeed becomes broader and becomes more asymmetric with an increasing weight on the low energy side. Interestingly, **Figure S 6**-c shows that these three characteristic evolutions of the $A_{1g}^{(2)}$ mode, namely softening, broadening and increasing asymmetry, are also observed at the composition-induced insulating (PI) to metal (PM) transition between $(V_{0.975}Cr_{0.025})_2O_3$ and $V_2O_3$. On the theoretical side, the anomalous asymmetric line shape has been assigned to a Fano resonance effect between the $A_{1g}$ phonons and electronic transitions within the $a_{1g}$ – bonding band.[10] According to this approach, the smaller asymmetry observed in the insulating phases results from a lower population of the $a_{1g}$ electronic band.[13]

Interestingly, the study carried out in ref. 10 has demonstrated that the asymmetry effect is also very strong for the $A_{1g}^{(1)}$ phonon located around 240 cm$^{-1}$. **Figure S 7**-a-b shows that the Raman spectrum between 160 and 290 cm$^{-1}$ can be decomposed as the sum of two phonons modes, a symmetrical $E_g^{(1)}$ mode and an asymmetrical $A_{1g}^{(1)}$ mode.

The same decomposition is also able to describe the Raman spectra on both sides of the pressure-induced insulator to metal transition, as shown in **Figure S 7**-d. The comparison between the electric–field-induced and pressure-induced transitions clearly shows a similar evolution of the asymmetry and linewidth parameters. The only difference between these two samples is the frequency of the $A_{1g}^{(1)}$ phonon mode, 253 cm$^{-1}$ in the crystal used for pressure experiments against 245 cm$^{-1}$ for the one used under electric field. **Figure S 7**-e shows that this small difference results from slightly different composition of these two crystals, $x_{Cr}$ = 3.8 % and 2.5 %, respectively.

Overall, these analyses of Raman spectroscopy in $(V_{1-x}Cr_x)_2O_3$ single crystals demonstrate that the electric-pulse-induced conducting filament (i) is composed only of $(V_{1-x}Cr_x)_2O_3$, without any other phase with different chemical composition, and (ii) presents vibrational signature similar with those observed at the pressure- and composition-induced insulator to metal transitions.

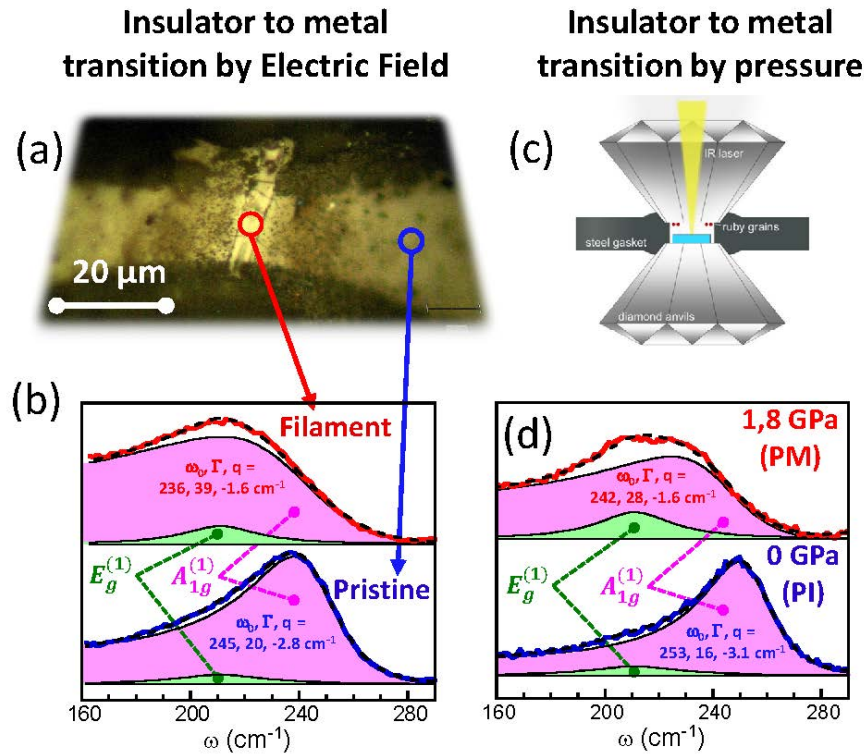

**Figure S 7** : (a-b) same Raman data as in **Figure S 6**, but zoomed on the low energy part of the spectra.

About the issue of non-stoichiometry in $V_2O_3$

We propose here a discussion about the question of non-stoichiometry in the $V_2O_3$ system in regards of the current development of OxRAM memories. First, we recall that OxRAM memories are based on insulating oxides in which the creation of oxygen vacancies leads to a local metallic state by changing the valence of the transition metal. These oxygen vacancies can diffuse and reorganize themselves in filamentary form by the application of an electric field. The presence of such a mechanism in many complex oxides has led to the misconception that oxygen *vacancies* are present in *all* complex oxides. However, there are two classes of oxides, on the one hand those where the creation of an oxygen vacancy is energetically favorable and on the other hand those where the most stable vacancies are located at the cation site. An analysis of the literature clearly shows that $V_2O_3$ belongs to this second category.

It was indeed both shown and explained that the exact formulation is $V_{2-y}O_3$, and NOT $V_2O_{3-x}$ or $V_2O_{3+x}$, x>0). All the direct compositional studies of $V_2O_3$ (see *e.g.* Refs. [14, 15]) actually conclude that non-stoichiometry, at least at the average macroscopic level, exists only on the "oxygen-rich" side. This result, based on very reliable mass gain during the re-oxidation of $V_2O_3$ to $V_2O_5$, proves that the real average composition is either $V_{2-y}O_3$ or $V_2O_{3+x}$. Let us recall that the corundum structure of $V_2O_3$ is based on a hexagonal close packing arrangement of oxygen ions, for which there is simply no space at all for any additional oxygen interstitial site. It rules out the formulation $V_2O_{3+x}$ and lets the formulation $V_{2-y}O_3$ as the only candidate. In addition, no direct study has shown the existence of a composition with oxygen vacancies, *i.e.* with a formula $V_2O_{3-x}$. From a chemical point of view, this is quite easy to understand: $V_{2-y}O_3$ (or $V_2O_{3+x}$) lead to the creation of $V^{4+}$ which ionic radius (0.72 Å) is very close to the one of $V^{3+}$ (0.78 Å) existing in stoichiometric $V_2O_3$. Conversely, the formulation with oxygen vacancies ($V_2O_{3-x}$) implies the presence of $V^{2+}$, with a huge ionic radius (0.92 Å) hardly compatible with the size of the octahedral $VO_6$ site present in $V_2O_3$. This statement is in line with recent theoretical calculations indicating that oxygen **vacancies** in $V_2O_3$ are energetically strongly disfavored with respect to vanadium vacancies.[16]

The recently published theoretical analysis on vacancies in $V_2O_3$ (see Ref.[16])) moreover demonstrates for the first time that the presence of **oxygen vacancies would lead to an *increase* of the AFI-PM transition temperature $T_{AFI-PM}$**., while vanadium vacancies tend to strongly decrease $T_{AFI-PM}$. As shown in **Figure S 5** (see also similar results obtained in Ref. [17]), it is obvious that new transitions that appears in the conducting filament occur at temperature lower than that of unaffected regions. It seems hence to discard the scenario of a conducting filament mainly composed of $V_2O_3$ with oxygen vacancies.

In addition, the scenario of vanadium vacancies within the filament can be discarded based on the evolution of the *a* and *c* cell parameters. As demonstrated in many early works,[7,14,18] the presence of vacancies on the vanadium site in non-stoichiometric $V_2O_3$ leads to a usual steric effect, with a simple decrease of both the *a* and *c* parameters. Conversely, the µ-XRD study of the electric-pulse-induced filament in $(V,Cr)_2O_3$ reveals a strikingly different behavior with a decrease of *a*, but an **increase** of the *c* unit cell parameter. In contrast, this unusual evolution of the cell parameters is strikingly similar to the one observed at the pressure-induced insulator to metal transition in (V,Cr)$_2$O$_3$ shown in **Figure S 2**. This key observation of the cell parameter change within the conducting filament supports the scenario proposed in our article, *i.e.* a conducting filament made of compressively strained (V,Cr)$_2$O$_3$.